\documentclass[12pt]{article} 
\usepackage{graphicx}
\title{Influence of chemical reactions on the nonlinear dynamics of dissipative flows}
\author{A.R. Karimov\\
\thanks{E-mail:alexanderkarimov999@gmail.com, arkarimov@mephi.ru}
Institute for High Temperatures, \\
Russian Academy of Sciences,\\
Izhorskaya 13/19, Moscow 127412, Russia \\
and Department of Electrophysical Facilities, \\ National Research Nuclear University MEPhI, \\
Kashirskoye shosse 31, Moscow, 115409, Russia
\\ A.M. Korshunov\\
Department of Electrophysical Facilities, \\ National Research Nuclear University MEPhI, \\
Kashirskoye shosse 31, Moscow, 115409, Russia
\\ V.V. Beklemishev\\
Department of Electrophysical Facilities, \\ National Research Nuclear University MEPhI, \\
Kashirskoye shosse 31, Moscow, 115409, Russia}
\date{ }
\begin{document} 
\maketitle
\newcounter{graf}
\begin{abstract}
The nonlinear dynamics of resistive flow with a chemical reaction is studied. Proceeding from the Lagrangian description, the influence of a chemical reaction on the development of fluid singularities is considered. \end{abstract}

\section{Introduction}

The clustering of matter in different physical conditions remains a topic of interest in modern science because of its importance for understanding the behavior of natural and artificial systems far from equilibrium \cite{lev}-\cite{kur}. The process of clustering typically develops in particle ensembles with long-range forces, such as gravity or electric interactions and it may lead to the formation of stable spatially inhomogeneous structures \cite{zel} - \cite{dn}. However, there is another possibility of structure formation in dissipative systems, and typical example of such systems is flow with some chemical reactions. 
In this paper, we are going to examine some special features of the nonlinear dynamics of such flows.

The macroscopic dynamics of dissipative systems with chemical reactions is, as a rule, described by diffusion equations \cite{lev}-\cite{nic}. However, such description is valid only for the systems with microscopic flow-transfer near some equilibrium state. If it is not the case, and we wish to consider the flow far from equilibrium, such approach is inadequate since the dynamics can involve several time and space scales that usually differ by several orders of magnitude. As a result, a full nonlinear treatment on the basis of hydrodynamic equations with chemical sources will be required to investigate the evolution of such system. 

Here we consider a particularly simple form of nonlinear dynamics of dissipative, one-dimensional flow moving in an external field with model chemical terms. We present a class of exact solutions of the fully nonlinear hydrodynamic equations describing the flow with a chemical reaction which may be useful to define the direction of the real system behavior. Based on this description, we compare the formation of density profile in the dissipative flows with a chemical reaction with the corresponding evolution in the flows without chemical reactions.

\section{Model}

We shall study the one-dimensional, time-dependent pattern for the infinite medium containing movable, particles of type $A$ and immobile particles of the $B$ type between which there is a chemical reaction leading to the formation of new immobile component $AB$. Such situation can be realized when $m_A \ll m_B$, where $m_A$ and $m_B$ are the mass of $A$ and $B$ particles, respectively. It is assumed that the movement of the component $A$ is caused by an external force (for definiteness, let this be the gravitational field and all particles are the neutral ones) and friction force (Stokes force) only. The macroscopic model of these processes can be present in an uniform form as
\begin{equation} 
{\partial n_A \over \partial t} + {\partial \over \partial x}(n_A u) = -W\/,
\label{1_chem}
\end{equation}
\begin{equation} 
{\partial n_B \over \partial t} = -W\/,
\label{1a_chem}
\end{equation}
\begin{equation} 
{\partial n_{AB} \over \partial t} = W\/,
\label{1b_chem}
\end{equation}
\begin{equation} 
{\partial u \over \partial t} + u {\partial u \over \partial x} = -\left(\nu + {W\over n_A}\right) u + g\/, 
\label{2_chem}
\end{equation}
where $u$ is the velocity of the $A$ component, $n_A$, $n_B$ and $n_{AB}$ is the density of the $A$, $B$ and $AB$ components, respectively, $\nu$ is the collision frequency, $g$ is the acceleration of freedom fall (note that the direction of the $0x$ axis is assumed to coincide with the direction of the acceleration of freedom fall $g$) and $W$ is the rate of the chemical reaction in the system. The concrete form of $W$ is defined by the order of reaction with respect to the components $A$ and $B$. Here we take for consideration the reactions of first ($s=1$) and second $(s=2)$ order only, and in this case we have
\[ W = \left\{\begin{array}{ll}
k_1 n_A,  & s=1, \\
k_2 n_A n_B,  & s=2,
\end{array}
\right.
\]
where $k_s$ is the chemical reaction constant. For simplicity, here we neglect the dependence of the friction coefficient $\nu$ on the medium parameters, as well as the dependence of the reaction constants $k_s$, setting $k_s$=const. and $\nu$=const. since these dependences are not important for our study of qualitative properties of the flow dynamics. 

\section{Lagrangian frame}

In order to see how these factors manifest themselves in the system dynamics it is convenient to pass from the Euler description of the original system (\ref{1_chem})-(\ref{2_chem}) to the Lagrangian frame. According to the definition (see, for example, \cite{sch_2, dav})
\begin{equation} 
\tau=t,\hspace{11mm} \xi=x-\int_0^t u(\xi,t^{\prime})dt^{\prime}\/,
\label{4_chem}
\end{equation}
where $x(\xi, t)$ satisfies the initial condition
$$x( \xi, 0)=\xi$$
and provides
\begin{equation} 
u(\xi, \tau) =\left({\partial x \over \partial 
\tau} \right)_{\xi}\/,
\label{U_nature} 
\end{equation}
the temporal and spatial derivatives are transformed as
\begin{equation} 
{\partial \over \partial t} \to {\partial  \over \partial \tau} - {u \over J} 
{\partial  \over \partial \xi}, 
\hspace*{11mm}
{\partial \over \partial x} \to {1 \over J} 
{\partial \over \partial \xi}
\label{6_chem}
\end{equation} 
with Jacobian
\begin{equation}
J(\tau, \xi ) = {\partial x\over \partial \xi} > 0\/, 
\label{5_chem}
\end{equation} 
besides, the condition of sign conservation of $J( \xi,\tau)$ eliminates singularities of the flow. 

Under the transformation (\ref{6_chem}), the Eq.(\ref{2_chem}) is reduced to 
\begin{equation} 
{\partial u \over \partial \tau} + \gamma u = g\/,
\label{9_chem}
\end{equation}
where 
\begin{equation} 
\gamma= W/n_A +\nu\/.
\label{g_chem}
\end{equation}
At the same time the continuity equations (\ref{1_chem}) is transformed into
\begin{equation} 
{\partial \over \partial \tau}\ln(J n_A)=-{W \over n_A}\/.
\label{14_chem}
\end{equation}
In such form this equation  will be needed for the exposition in the next sections.
\section{Dynamics for $s=1$}

Following \cite{chm}, we start from the flow with a chemical reaction of first order. In fact, if $\gamma$=const., then Eq. (\ref{9_chem}) is the Newton equation for a particle, moving under the influence of gravity and friction. Such situation occurs when the chemical reaction is first order, i.e. $s=1$. Physically this means that the concentration of one component is excessive, for example, in our model $n_A \ll n_B$. In this case the solution of Eq. (\ref{9_chem}) has the exact solution
\begin{equation} 
u(\xi,\tau) = u_0(\xi) e^{-\gamma \tau} + {g \over \gamma}
\left(1 - e^{-\gamma \tau}\right)\/,
\label{10_chem}
\end{equation}
where $u_0(\xi)$ is initial velocity and $\gamma= k_1 +\nu$. With the help of the relation (\ref{U_nature}) bearing in mind $x(\xi, \tau=0) = \xi$, we may deduce the path of a fluid element 
\begin{equation} 
x(\xi,\tau) = \xi + {g\over \gamma}\tau + {1\over \gamma} 
\left( u_0 - {g \over \gamma}\right)(1 - e^{-\gamma \tau})\/,
\label{12_chem}
\end{equation}
from which we find 
\begin{equation}
J(\xi,\tau) = 1 + {u_0^{\prime}\over \gamma}(1 - e^{-\gamma \tau})\/.
\label{J_chem}
\end{equation}
Finally, substitution of this relation and $W$ for $s=1$ into (\ref{14_chem}) leads to 
\begin{equation} 
n_A(\xi, \tau) = {\gamma n_{0A} e^{-k_1 \tau}\over \gamma + (1 - e^{-\gamma \tau}) u_0^{\prime}}\/,
\label{na_chem}
\end{equation}
where $n_{0A} = n(\xi, \tau=0)$. Also, from relation
$${\partial u\over \partial x} = {1 \over J} {\partial u \over \partial \xi}$$
taking in account (\ref{10_chem}) and (\ref{J_chem}) we get
\begin{equation} 
{\partial u\over \partial x} = {\gamma u_0^{\prime} e^{-\gamma \tau}\over \gamma + (1 - e^{-\gamma \tau}) u_0^{\prime}}\/,
\label{u_chem}
\end{equation}

As is seen from (\ref{na_chem}) and (\ref{u_chem}), the chemical reaction affects the flow via the parameter $\gamma$. It is  worth noting that in the limit $\gamma \to 0$ the relations (\ref{na_chem}) and (\ref{u_chem}) pass to 
\begin{equation} 
n_A = {n_{0A} e^{-k_1 \tau}\over 1 +\tau u_0^{\prime}}, \hspace{11mm} {\partial u\over \partial x} = {u_0^{\prime} \over 1 + \tau u_0^{\prime}}\/.
\label{15b_chem}
\end{equation}
In absence of a chemical reaction ($W \to 0$) and/or  friction ($\nu \to 0$) these relations describe the flow by inertia in collisionless neutral gas. When $ u_0^{\prime} <0 $, it predicts a singular behavior for finite time. A discontinuous pattern is a well-known intrinsic feature of fluids especially in inviscid pressureless limit. As is seen from (\ref{12_chem}), on times $\tau < \gamma^{-1}$ the motion of fluid elements occurs in a similar way. 

However, in our model of dissipative flow with chemical a reaction some new peculiarities appear. In the case when $\gamma \neq 0$ and $W \neq 0$ the collapse condition is 
\begin{equation} 
1 + {1 - e^{-\gamma \tau}\over \gamma} u_0^{\prime} = 0
\label{16_chem}
\end{equation}
from which follows the collapse time
$$\tau_*= -{1\over \gamma }\ln\left(1 + {\gamma \over u_0^{\prime}}\right)\/.$$
If $\gamma \to 0$, this estimation transforms into $\tau_* = -1/ u_0^{\prime}$ for the Eq. (\ref{15b_chem}). It is clear that $\tau_*$ for the dissipative flow with a chemical reaction is larger than the corresponding values for the collisionless flow. Thus, the chemical reaction and the friction add some restriction on the parameters of the problem but do not limit the development of singularity in the system.

The physically admissible parameters $u_0$ and $\gamma$ are determined by the condition that the density should be positive for all time, i.e. the following inequality must be fulfilled 
\begin{equation} 
u_0^{\prime} <0, \hspace{11mm} \mid u_0^{\prime} \mid \geq  \mid \gamma \mid\/.
\label{17_chem}
\end{equation}
It should be noted that in this simplest case the nonlinear dynamics for some initial conditions can lead to the formation of dynamical space structures \cite{ks,kss} in which the initial conditions play the role of a driving parameter precluding the formation of singularities but providing for the formation of transient dynamical structures.

\section{Dynamics for $s=2$}

Now we move on to a more realistic case of $s=2$. For simplicity, in this section we shall restrict our consideration to the limit $W/n_A \ll \nu$ when we may use the relation (\ref{J_chem}) for our estimations. The Eq. (\ref{14_chem}) then reduces to 
\begin{equation} 
{\partial \over \partial \tau}\ln(J n_A)=- k_2 n_B\/.
\label{18_chem}
\end{equation}
Differentiating relation (\ref{18_chem}) with respect to $\tau$ and using (\ref{1a_chem}), we get
\begin{equation} 
{\partial^2 \over \partial \tau^2}\ln(J n_A)= k_2^2 n_B n_A\/.
\label{19_chem}
\end{equation}
Then eliminating $n_B$ from (\ref{19_chem}) with the help of (\ref{18_chem}) we obtain
\begin{equation} 
J{\partial^2 y\over \partial \tau^2} + k_2 {\partial e^y \over \partial \tau}=0\/,
\label{20_chem}
\end{equation}
where we introduce
\begin{equation} 
y = \ln(J n_A)\/.
\label{21_chem}
\end{equation}
It should be noted that in the framework of model worked out the Eq. (\ref{20_chem}) is an exact equation, one that has been correct for any $J$.

Now we have to define the initial conditions for Eq. (\ref{20_chem}) which follow from the initial conditions for $n_A$, $n_B$ and $J$. Without losing too much generality, we can set $n_{0A}=1$ and $n_{0B}=\varepsilon<1$ so we have $y(\tau=0)=0$. Proceeding from obvious relations
$${\partial y\over \partial \tau}= {J\dot{n}_A+n_A\dot{J} \over J n_A}, \hspace{7mm} {\partial J\over \partial \tau} = u_0^{\prime}\/,$$
where the overhead dot now denotes derivative with respect to $\tau$, and Eq. (\ref{18_chem}) rewritten for $\tau=0$ in the form 
$$\left.{\partial n_A\over \partial \tau}\right|_{\tau=0}=-n_{0A} u_0^{\prime} - k_2n_{0A}n_{0B}$$
we get
$$\left.{\partial y\over \partial \tau}\right|_{\tau=0}=-k_2\varepsilon\/.$$
Thus, the Eq. (\ref{20_chem}) together with the initial conditions
\begin{equation} 
y(\tau=0)=0, \hspace{11mm} \left.{\partial y\over \partial \tau}\right|_{\tau=0}=-k_2\varepsilon\/.
\label{22_chem}
\end{equation}
defines the evolution of the flow with a chemical reaction of second order. 

Unfortunately, this equation cannot be solved analytically but proceeding from a Chaplygin comparison theorem for nonlinear differential equations \cite{bel} we are able to obtain the apriori estimates for the solutions of problem (\ref{20_chem})-(\ref{22_chem}). Applying this theorem to our issue, the functions $y_{min}(\tau)$ and $y_{max}(\tau)$ satisfying the initial conditions (\ref{22_chem}) shall hold the relation 
$$y_{min} \leq y \leq y_{max}$$ 
in the interval $0 \leq \tau \leq \tau_*$ if the following inequalities are fulfilled 
$$\Lambda(y_{min}) <0, \hspace{11mm} \Lambda(y_{max}) >0$$ 
where
\begin{equation} 
\Lambda(y) = J{\partial^2 y\over \partial \tau^2} + k_2 {\partial y \over \partial \tau} e^y\/.
\label{23_chem}
\end{equation} 

As such functions we can take 
\begin{equation} 
y_{min}= - k_2 \varepsilon \tau, \hspace{11mm} y_{max}={a \over 2}\tau^2 - k_2 \varepsilon \tau \/,
\label{24_chem}
\end{equation}
where $a$ is some unknown parameter to be determined. Indeed, for $y_{min}$ we get 
$$\Lambda(y_{min}) = -k_2^2\varepsilon \exp(-k_2\varepsilon \tau) <0\/,$$ which means that we can consider $y_{min}$ as a  lower boundary of the exact solution for Eq. (\ref{20_chem}). We now look for the parameter $a$ for which $y_{max}$ becomes an upper boundary of the exact solution Eq. (\ref{20_chem}). Substitution of $y_{max}$ into (\ref{23_chem}) yields 
\begin{equation} 
\Lambda(y_{max}) = (1 + \tau u_0^{\prime}) a - k_2(k_2\varepsilon - a \tau) \exp\left({a\over 2}\tau^2 - k_2\varepsilon \tau\right)\/.
\label{25_chem}
\end{equation}
If we set
\begin{equation} 
a > k_2^2\varepsilon, \hspace{11mm} a > k_2\varepsilon \mid u_0^{\prime} \mid\/.
\label{26_chem}
\end{equation}
then the curve $y_1(\tau) = k_2^2\varepsilon - k_2a \tau$ always is lower than the curve $y_2(\tau) = a + u_0^{\prime}a\tau$ in the interval $\tau \geq 0$. Moreover, it easy to see that 
$$ k_2(k_2\varepsilon - a \tau) \exp\left({a\over 2}\tau^2 - k_2\varepsilon \tau\right) \leq k_2(k_2\varepsilon - a \tau)$$
in the interval $0 \leq \tau \leq k_2\varepsilon /a$. For $\tau \geq k_2\varepsilon /a$ the second term of (\ref{25_chem}) becomes a positive value and the condition $\Lambda(y_{max})>0$ always holds true if the parameter $a$ belongs to the interval defined by the relation (\ref{26_chem}). It implies that the functions $y_{min}(\tau)$ and $y_{max}(\tau)$ have no peculiarities in the interval $0 \leq \tau \leq \tau_*$. Then from the relation (\ref{21_chem}) it follows that in the present case there may be only one hydrodynamic peculiarity associated with $J = 0$ which is similar to the situation occurring in flow with  the reaction of first order. 
\begin{center}
\includegraphics[width=9.cm]{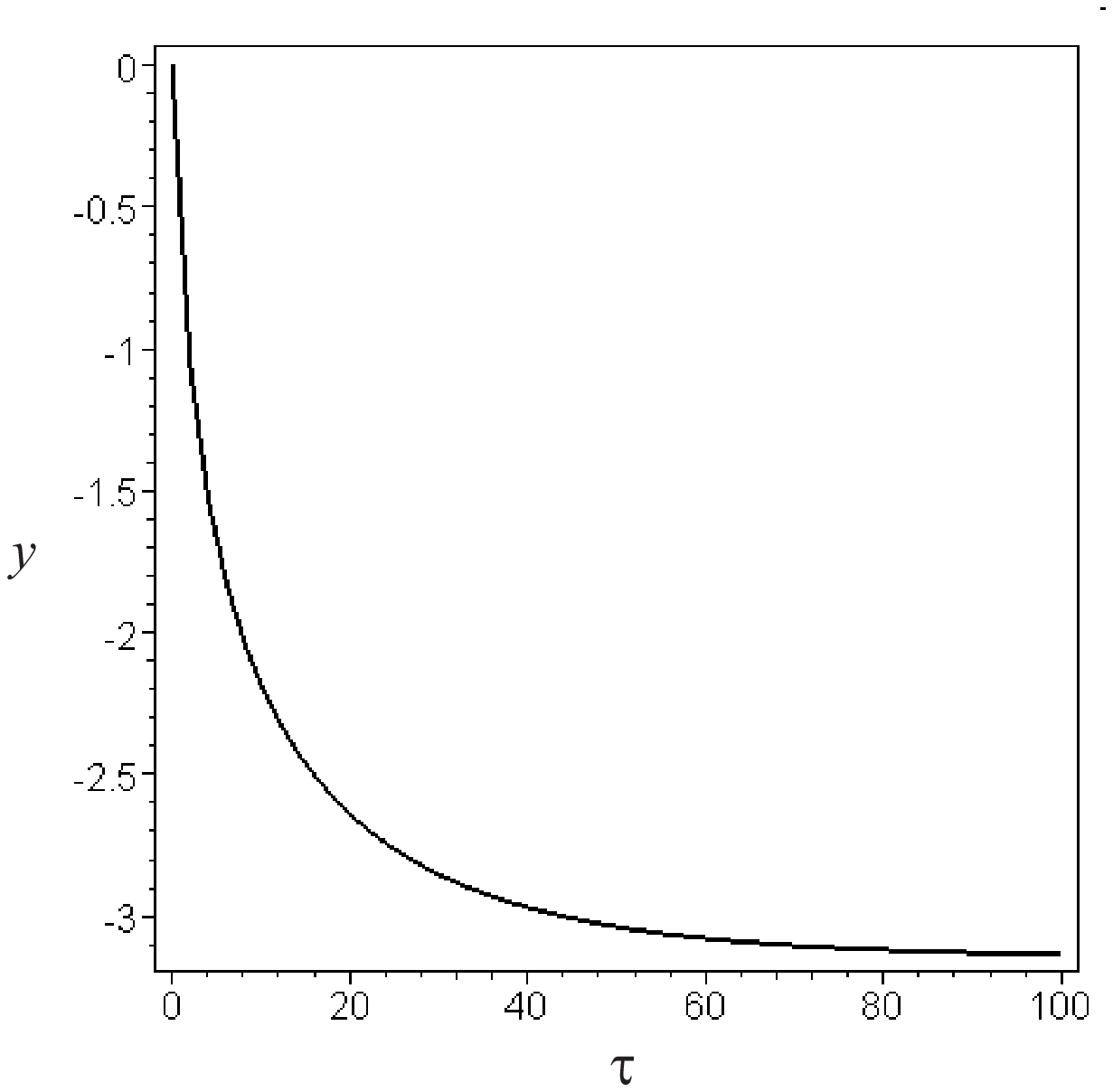}  
\end{center}
\hspace*{15mm}\parbox[b]{14cm}{\normalsize Fig.1. \hspace*{1mm} Time dependences of $y(\tau)$ for $k_2 =1$, $u_0^{\prime}=-10^{-3}$ and $k_2\varepsilon=-0.96$.}\\\\

As an illustration of exact dynamics of the flow for case $s=2$, in Fig.1 we present the numerical solution of (\ref{20_chem}) for $k_2 =1$, $u_0^{\prime}=-10^{-3}$ and $k_2\varepsilon=-0.96$. This solution exhibits regular behavior up to the moment of development of hydrodynamic collapse $\tau_*$ and the function $y(\tau)$ holds negative in the interval $0 \leq \tau < \tau_*$. Thus, this partial case indicates that our conclusion about character of $y(\tau)$ remains valid and collapse can be caused only by the flow hydrodynamics. 
\begin{center}
\includegraphics[width=9.cm]{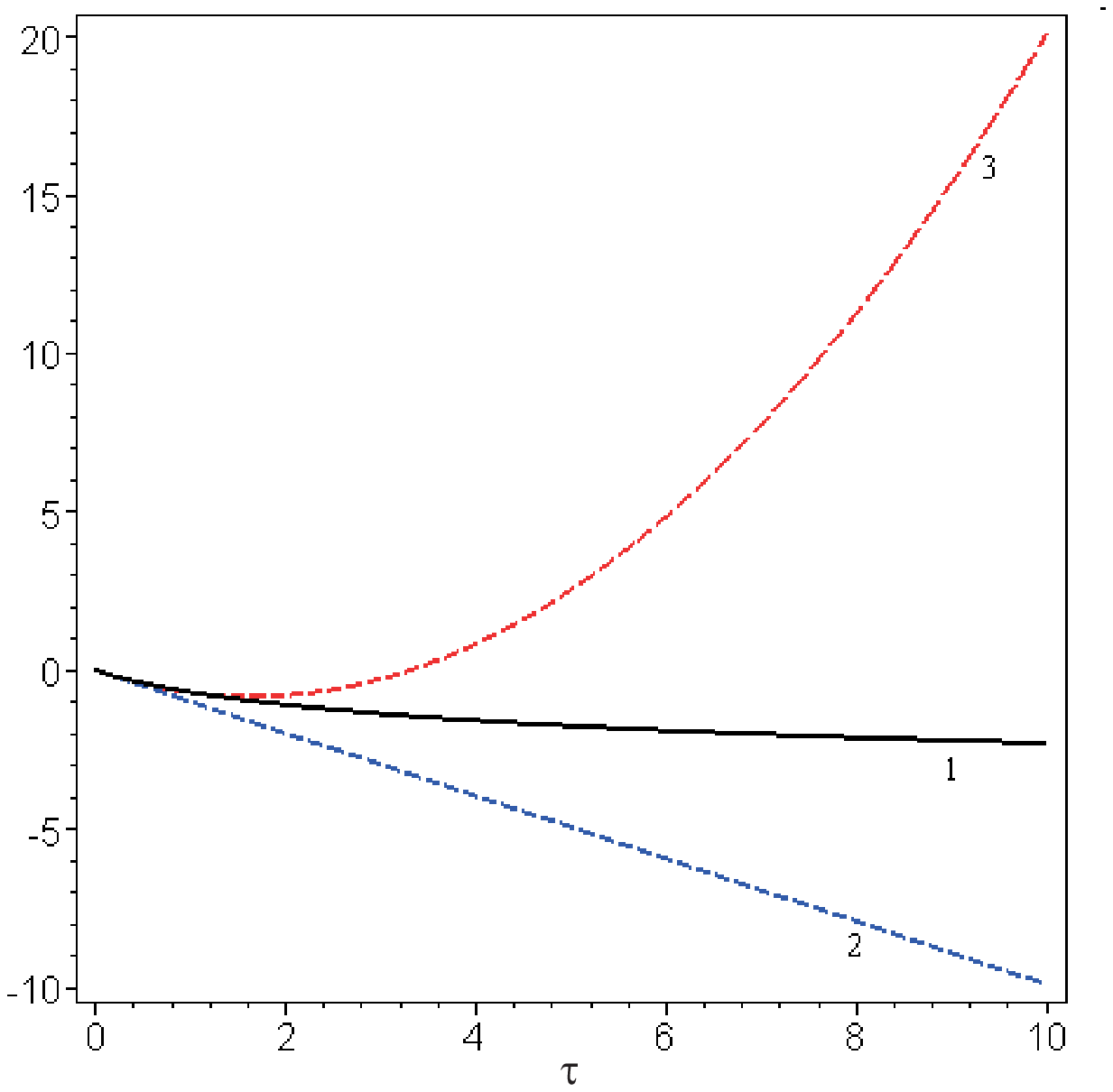}  
\end{center}
\hspace*{15mm}\parbox[b]{14cm}{\normalsize Fig.2. \hspace*{1mm} Time dependence of $y$ (line 1), $y_{min}$(line 2) and $y_{max}$ (line 3).}\\\\

In Fig.2 we graph this exact solution with the lower $y_{min}$ and upper $y_{max}$ boundaries for small times. As is seen from these graphs, we have a good approximation of the exact solution near initial state but for large times the boundaries can point out some rough estimate of the exact solution. However, such method of estimates seems sufficient enough for the study of the effect of chemical reactions on the formation of singularities and the influence of initial conditions on the dynamics of the system worked out.

\section{Conclusion}

In this paper we have focused on the influence of a chemical reaction on the dynamics of the dissipative, one-dimensional flow. In order to get a full analytical description we have considered the reaction of first and second orders. These reactions belong to the simplest type of possible reactions. However, the present results indicate that the same features can be observed in more complex systems. 

In particular, we have shown that the wave breaking may occur in the flow under analysis [see Eq. (\ref{14_chem})] similarly to what happens in a collisionless flow [see Eq. (\ref{15b_chem})]. In the present example the dissipative physical processes changed only the kind of singular dynamics but did not eliminate the phenomenon itself. It is important to stress that the above-described behavior can be observed only for the initial conditions and the parameters stated in Eq. (\ref{17_chem}). However, in the present case the evolution of the resistive flow is not restricted by any physical mechanism, such as the pressure gradient, which is usually presumed to limit the growth of the density peak (see, for example, \cite{sch_2,k02}).

Such type of solution represents a collapse-like class of nonlinear solutions that may arise in different physical situations. In particular, the results describing the formation of time-dependent structures in dissipative flow may be of interest for some biophysical experiments in laboratory conditions \cite{es,ep,van} and aerosol applications \cite{reist,smir}. Besides, we believe that our present approach can be generalized, for instance, to an inhomogeneous, many-component fluid or plasma flows in gravitational or electrostatic fields \cite{kur,sch_2,ep,smir,ss09,kys12,inh}. 

It should thus be of interest to examine such flows in the cylindrical and spherical geometries, which have more natural rotational degrees of freedom (see, for example, \cite{es,sch_2,s96}). Owing to different kinds of interactions which may exist in such multidimensional systems and due to a variety of their initial states, different modes of collective motion are possible \cite{ss09}-\cite{k09}. Therefore, in higher spatial dimensions, a large variety of nonlinear dynamical structures can be expected \cite{sch_2,nat}. However, one should be noted that depending on the ratio between nonlinearity and dispersion, one can expect the formation of hydrodynamic collapses or nonlinear wave structures (see, for example, \cite{dav,nat,ir}). In our case, the present results for reactions of first and second orders (in the limit $W/n_A \ll \nu$) indicate that there is a relatively weak effect of chemical terms on the flow dynamics since the nonlinearity is too strong that the system experiences a collapse-like behavior for small times. Proceeding from this point one may expect that singularities form cellular structures in multidimensional geometry. As a result, the intensity of the chemical reactions strongly can increase at these points. Such special features are expected to play an important role, in particular, for the understanding of basic properties of aerosol systems in various environments and laboratory conditions \cite{smir,smir_2}. But it is only our assumption or guess and no more. In order to show the realization of this hypothetical mechanism we should study the above-outlined script of dynamics for dissipative flows with a chemical reaction in multidimensional geometry. 

\end{document}